%% file: main.tex
\documentclass[11pt]{article}

\usepackage[T1]{fontenc}
\usepackage{lmodern}
\usepackage{authblk}

\usepackage[left=2.2cm,right=2.2cm,top=2.4cm,bottom=2.4cm]{geometry}
\usepackage{graphicx}
\usepackage{amsmath}
\usepackage{amssymb}
\usepackage{booktabs}
\usepackage{natbib}      %
\usepackage{lineno}      %
\usepackage{setspace}
\usepackage[hidelinks]{hyperref}
\usepackage{placeins}

\input{notation}

\title{\textbf{Silscore: modelling siRNA potency with\\
positional thermodynamics and data-selected\\
nucleotide-identity terms}}
\author[1]{Giuseppe Sacco\thanks{Co-corresponding authors: Giuseppe Sacco
(\href{mailto:gsacco@sissa.it}{gsacco@sissa.it}) and Giovanni Bussi
(\href{mailto:bussi@sissa.it}{bussi@sissa.it}).}}
\author[2]{Isabella Baldini}
\author[2]{Riccardo Paone}
\author[3,2]{\authorcr Mattia Capulli}
\author[1]{Guido Sanguinetti}
\author[1]{Giovanni Bussi\protect\footnotemark[1]}
\affil[1]{SISSA -- International School for Advanced Studies, Trieste, Italy.}
\affil[2]{Silthera, L'Aquila, Italy.}
\affil[3]{Department of Biotechnological and Applied Clinical Sciences, University of L'Aquila, Italy.}
\date{November 11, 2026}

\begin{document}

\maketitle

\section*{Abstract}
Small interfering RNAs (siRNAs) provide a programmable route to gene silencing, but candidates
targeting the same transcript can differ markedly in potency. Existing predictors range from
interpretable empirical rules with limited accuracy to high-capacity models that can lose
performance when datasets or targets change. To address this, we present Silscore, a guide-only
model that combines a quadratic preference for overall duplex stability, a position-resolved
profile of nearest-neighbour stacking energies and
sparse nucleotide-identity terms selected from residuals left unexplained by thermodynamics. Its
main fit uses 26 parameters, fewer than half as many as any trainable comparator evaluated here.
We evaluate Silscore on the canonical split of the Huesken dataset, two external datasets
(Mixset and Katoh--Suzuki), and an evaluation in
which the same guides are tested after training on dissimilar targets. Silscore attains the highest
observed Pearson correlation in all three transfer evaluations, and the data support an advantage
over every applicable comparator on Katoh--Suzuki. For a fixed evaluation set, we compare two
equally sized training sets: one with sequences more similar to the evaluation sequences and one
with less similar sequences. After training on the less similar sequences,
Silscore has the highest observed correlation, and the data do not establish a change in its performance. The fitted parameters
can also be interpreted directly: identity terms locate
sequence effects that base-pairing energies cannot express, global stability has a data-estimated
intermediate optimum, and local energy profiles recover terminal asymmetry and a preference for
weaker seed and stronger central pairing. Silscore thus provides a compact, data-driven model that
combines strong predictive performance with robustness when tested in new settings and expresses
previously reported empirical preferences as quantities estimated from data.

\input{sections/introduction}

\input{sections/results}
\input{sections/discussion}

\input{sections/methods}

\input{sections/declarations}

\bibliography{refs}

\clearpage
\input{sections/supplementary}

\end{document}

%% file: notation.tex
\newcommand{\dG}{\Delta G}

%% file: sections/introduction.tex
\section{Introduction}\label{sec:introduction}

Selective gene silencing is a powerful way to perturb gene function and to suppress
disease-causing transcripts, including those involved in genetic disorders
\citep{conroyChemicalEngineeringTherapeutic2022}. Small interfering RNAs (siRNAs) provide such a
sequence-programmable route \citep{patrickwaltonDesigningHighlyActive2010}, yet
candidates can differ markedly in potency, making efficacy prediction central to siRNA
design \citep{hueskenDesignGenomewideSiRNA2005}. This sequence dependence reflects how the
silencing machinery processes the siRNA. The duplex is loaded into Argonaute, the guide-binding
protein at the core of this machinery, and one strand is retained as the guide. Strand choice
depends in part on the relative stability of the two ends of the siRNA duplex
\citep{khvorovaFunctionalSiRNAsMiRNAs2003,schwarzAsymmetryAssemblyRNAi2003}; once loaded,
different guide regions support distinct binding and catalytic roles
\citep{weeArgonauteDividesIts2012}.
Thermodynamics has been part of siRNA design from the earliest empirical sequence and stability
rules \citep{reynoldsRationalSiRNADesign2004,ui-teiGuidelinesSelectionHighly2004}. A fitted
positional profile of nearest-neighbour stacking energies was less accurate than the sequence
representations compared with it \citep{vertAccurateInterpretableModel2006}, whereas other studies
excluded guides predicted to form excessively stable duplexes or identified an intermediate range of overall duplex stability
\citep{ichiharaThermodynamicInstabilitySiRNA2007,matveevaOptimizationDuplexStability2010}.
Recent deep-learning models combine thermodynamic descriptors with sequence and target-context
features \citep{longSiRNADiscoveryGraphNeural2024,baiOligoFormerAccurateRobust2024}.
Together, these studies indicate that thermodynamics is informative. However, some sequence effects cannot be expressed
through the base-pairing energies used here. A clear example is the $5'$ nucleotide of the guide,
which Argonaute recognises through a direct protein contact
\citep{frankStructuralBasis5nucleotide2010}. This motivates a hybrid representation in which a
thermodynamic profile provides the backbone and a few nucleotide-identity corrections capture what it leaves
unexplained. Transfer must also be tested explicitly: flexible models with many parameters can
learn experimental-design artefacts, while train--test overlap can inflate performance relative
to genuinely unseen targets
\citep{matveevaComparisonApproachesRational2007,liuMultitaskLearningCrossplatform2010,
caoInflatedPerformanceMeasures2019,longSiRNADiscoveryGraphNeural2024}.

Here, we present Silscore, a guide-only model of siRNA potency that combines a quadratic
preference for overall duplex stability, a constrained positional profile of nearest-neighbour
stacking energies and sparse nucleotide-identity terms. These are fitted corrections associated
with a specified nucleotide at a specified guide position and capture residual effects left
unexplained by the thermodynamic representation.
Trained on the Huesken partition, its main fit has 26 parameters,
fewer than half as many as any trainable comparator evaluated here and orders of magnitude fewer
than recent deep-learning architectures. We
evaluate it on the published held-out set and on the external Mixset and Katoh--Suzuki
datasets, which enter neither model fitting nor term selection in the performance evaluation
\citep{hueskenDesignGenomewideSiRNA2005,katohSpecificResiduesEvery2007,
baiOligoFormerAccurateRobust2024}. Although selected on Huesken alone, the identity terms improve
transfer over thermodynamics alone on both external datasets. Under a common reproducible protocol,
Silscore achieves the highest observed correlation in all three transfer evaluations, and the data
support an advantage over every applicable comparator on Katoh--Suzuki.
When the same guides are used for evaluation, the data do not establish a performance loss for
Silscore after the training set changes to guides against dissimilar targets.
We then examine what each fitted component reveals. The global term quantifies both an intermediate
optimum in duplex stability and the decrease in predicted potency away from it; the local profile estimates how pairing
stability should be distributed along the guide; and the sparse
identity terms locate sequence effects that cannot be expressed through the base-pairing energies
used here. The fitted parameters therefore place previously reported sequence and stability
preferences in a single model that estimates them from data rather than prescribing them as
qualitative rules.
Finally, we examine the variances and covariances of the score components, and
use the full model to map nucleotide preferences along the guide.

%% file: sections/results.tex
\section{Results}\label{sec:results}

\subsection{A physical model of siRNA potency}\label{sec:model}

Silscore models siRNA potency using three guide-derived components: a quadratic preference
for total duplex stability, a position-resolved profile of nearest-neighbour stacking energies,
and sparse nucleotide-identity terms selected from residuals left unexplained by thermodynamics.
The first two define the thermodynamic component; the third captures sequence contributions that
cannot be expressed in terms of the base-pairing energies used here.
Because all inputs are derived from the guide sequence alone, target-site accessibility and
transcript-level context are outside the model by construction.

Let $g$ be a guide of length $L=19$ nucleotides and $\dG_i$ the stacking free energy at
$37^\circ$C associated with its $i$-th dinucleotide step. We use the nearest-neighbour model
\citep{tinocoImprovedEstimationSecondary1973} with the parameters of
\citet{xiaThermodynamicParametersExpanded1998}. The overall duplex free energy $\dG(g)$ is
obtained from the 18 stacking terms together with initiation and terminal corrections
(Section~\ref{sec:thermodynamic-calculations}).
We model the predicted efficacy as
\begin{equation}\label{eq:score}
\hat{y}(g) = \beta_0
 \;-\; \underbrace{\kappa\bigl(\dG(g)-\dG^{*}\bigr)^{2}}_{\text{global stability}}
 \;+\; \underbrace{\sum_{i=1}^{18} w_i\,\dG_i}_{\text{local stability}}
 \;+\; \underbrace{\sum_{p=1}^{19} \theta_p(g_p)}_{\text{nucleotide identity}} \; .
\end{equation}
For record $j$, the measured efficacy is $y_j=\hat y(g_j)+\varepsilon_j$, where
$\varepsilon_j$ collects measurement noise and contributions not represented by the model.

We now explain each term one at a time. $\beta_0$ is a free unpenalised intercept, fitted jointly
with the other coefficients. It sets the level of the prediction and makes the fitted residuals
mean-zero, but is not otherwise interpreted.
The quadratic term accounts for the effect of overall duplex stability on potency, a classical determinant in the
design literature: duplexes that are too stable or too unstable are both held to be less potent
than those of intermediate stability \citep{matveevaOptimizationDuplexStability2010}, a
preference the empirical rules encode as a window on GC content
\citep{reynoldsRationalSiRNADesign2004}. The model captures this by penalising deviations from
an optimal free energy $\dG^{*}$ with a penalty strength $\kappa$, both fitted on the data.
The weights $w_i$ say where local stability helps or hurts. They are constrained to sum to zero,
so that the local profile describes positional differences rather than absorbing part of the
global-stability contribution.
Fitting weights on a
position-resolved profile of stacking energies has already been investigated by
\citet{vertAccurateInterpretableModel2006} as one of several they compared.
The last term carries additional sparse position-dependent corrections: $\theta_p(b)$ is a
coefficient for base $b$ at position $p$. The sum spans all 19 nucleotide positions and uses
the same coefficient map for every guide; coefficients are set to zero except for the few
position--base combinations selected on the training data. A well-understood example is the $5'$
nucleotide, read by Argonaute
through a protein contact rather than through base pairing
\citep{frankStructuralBasis5nucleotide2010}, which no rearrangement of stacking
energies can account for. We select the non-zero coefficients on the training data alone, from
the discrepancies the thermodynamics leaves unexplained: the model is fitted without identity
terms, and we retain the position--base combinations whose residuals depart from zero beyond a
multiplicity-corrected threshold (Section~\ref{sec:identity-selection}).

The coefficients follow from a closed-form minimisation over a small number of parameters
(Section~\ref{sec:model-fitting}): 26 in total for the main model discussed here, including the
six selected identity terms.

\subsection{Analysed datasets}\label{sec:data}

We use three public datasets of measured potency (Table~\ref{tab:datasets}): the reporter
screen of \citet{hueskenDesignGenomewideSiRNA2005}, the Katoh--Suzuki (KS) single-target EGFP
dataset \citep{katohSpecificResiduesEvery2007}, and the Mixset released with OligoFormer by
\citet{baiOligoFormerAccurateRobust2024}. The latter contains 472 siRNAs drawn from six of the seven
studies described by its authors and combines data from heterogeneous assays
\citep{amarzguiouiAlgorithmSelectionFunctional2004,harborthSequenceChemicalStructural2003,
khvorovaFunctionalSiRNAsMiRNAs2003,reynoldsRationalSiRNADesign2004,
ui-teiGuidelinesSelectionHighly2004,vickersEfficientReductionTarget2003}.
The same authors call the
KS dataset the Takayuki dataset.
Huesken spans 34 reporter constructs, Katoh--Suzuki a single
EGFP transcript, and Mixset multiple transcripts and assay types. Evaluating a model trained on
Huesken against the other datasets therefore changes the target composition and, for Mixset, the
type of measurement.

We fit on Huesken alone, using the train/test division published with the original screen to permit comparison with previously published results (2182
training and 249 test siRNAs).
Although siRNAs are held out from training for performance evaluation, the published Huesken split remains an in-dataset evaluation: the experimental context is unchanged, and the original per-siRNA partition does not enforce sequence dissimilarity between training and test.
It therefore provides an optimistic estimate of performance relative to transfer across experimental contexts or genuinely unfamiliar sequences.
The related role of sequence familiarity is examined explicitly in Section~\ref{sec:targets}.
Mixset and Katoh--Suzuki enter neither the fit nor the model selection unless explicitly stated.

\begin{table}[!htbp]
\centering
\begin{tabular}{lccc}
\toprule
 & Huesken & Mixset & KS \\
\midrule
siRNAs           & 2431 & 472 & 702 \\
targets          & 34 reporter inserts & multiple mRNAs & EGFP mRNA \\
assay            & eYFP/eCFP reporter & heterogeneous & EGFP reporter \\
readout          & protein & mixed & protein \\
role here        & training, held-out test & cross-dataset & cross-dataset \\
\bottomrule
\end{tabular}
\caption{The datasets analysed in this work. Because the experimental readouts and their
normalisation differ, absolute activity values are not assumed to be calibrated across datasets. Guides are 19 nucleotides throughout.}
\label{tab:datasets}
\end{table}

\subsection{Performance of the Silscore model}\label{sec:performance}

We evaluate three model configurations: Thermo-only, which disables identity-term
selection (20 free coefficients); Thermo+$\mathrm{U1}$, which fixes $\mathrm{U1}$ as its sole identity term
(21); and Silscore, which applies the complete selection protocol.
For the main Huesken fit, Silscore retains six identity terms: $\mathrm{U1}$, $\mathrm{C1}$, $\mathrm{C2}$, $\mathrm{G14}$, $\mathrm{G18}$ and $\mathrm{A19}$, giving 26 parameters in total; their interpretation
is developed in Section~\ref{sec:identity}, after evaluating their predictive contribution.
We compare with four selected representations of the Designer of Small Interfering RNA (DSIR) method
\citep{vertAccurateInterpretableModel2006}, i-Score re-fitted on our training set
\citep{ichiharaThermodynamicInstabilitySiRNA2007}, and the rule-based scores of
\citet{reynoldsRationalSiRNADesign2004} and
\citet{ui-teiGuidelinesSelectionHighly2004}, as well as OligoFormer
\citep{baiOligoFormerAccurateRobust2024}, a recent deep-learning model that combines thermodynamic
features with embeddings of the guide and its target-sequence context from the pretrained RNA
foundation model RNA-FM \citep{chenInterpretableRNAFoundation2022}. Its prediction network has
about 1.4 million trainable parameters, excluding the pretrained RNA-FM model.
DSIR uses position-specific nucleotide indicators (sparse), short-motif counts (spectrum), or
both (composite); the numbers indicate the sequence lengths used.
For the main Huesken split, the DSIR fits retain 54, 72, 84 and 96 non-zero coefficients for
sparse-19, spectrum-19, composite-19/19 and composite-21/21, respectively, compared with 77 for
i-Score.
Within each evaluation, all applicable methods are evaluated on the same records, and all fitted
methods start from the same training pool. For OligoFormer, 20\% of each pool is reserved for internal validation;
its reported correlation is the mean across 25
training runs, combining five training--validation splits with five PyTorch seeds, which control
weight initialization, dropout and data shuffling. The other fitted methods use no random
initialization and require no separate validation set, so each is fitted once on the full training set.
Reynolds and Ui-Tei are fixed rules and require no training
(Supplementary Section~\ref{sec:comparators}).

\begin{figure}[!htbp]
\centering
\includegraphics[width=\textwidth]{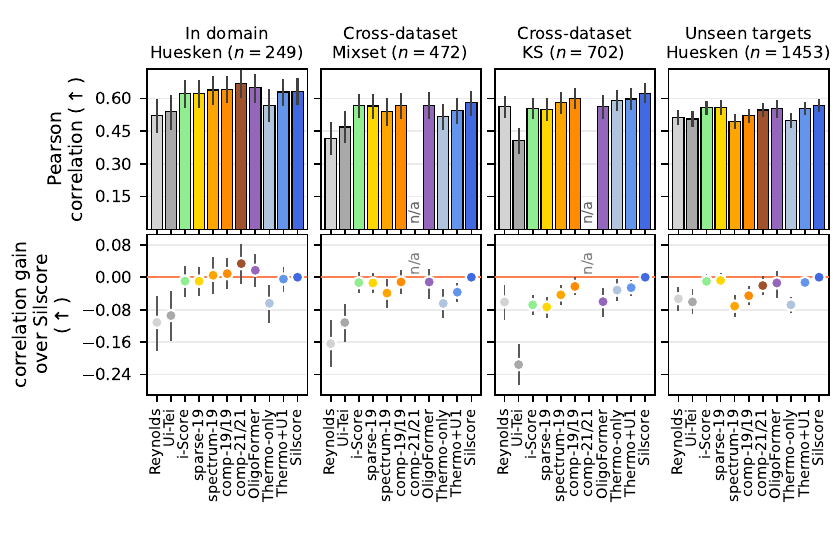}
\caption{Top: Pearson correlations in the four evaluations, with separate 95\% bootstrap
intervals. Bottom: paired differences in Pearson correlation between each method and Silscore
(method minus Silscore), with 95\% bootstrap intervals obtained by resampling the same records
for both methods in each of 4000 replicates; negative values favour Silscore. Sparse-19, spectrum-19,
composite-19/19 and composite-21/21 are DSIR representations; Thermo-only is the Silscore
variant in which identity-term selection is disabled. Composite-21/21 requires the $3'$
overhang and is therefore unavailable outside Huesken.
OligoFormer is represented by the mean correlation of 25 training runs; its intervals and
paired contrasts also resample training--validation splits and PyTorch seeds.
The fourth evaluation uses the origin of the target sequence to separate training and evaluation: all fitted models
are re-trained on 489 guides against rodent inserts and evaluated on 1453 guides against human
inserts, with all measurements made in the same human H1299 assay. Training Silscore on the rodent-insert guides
selects $\mathrm{U1}$ and $\mathrm{G18}$.}
\label{fig:performance}
\end{figure}

Figure~\ref{fig:performance} covers four evaluations. The first three---the held-out Huesken test
set and cross-dataset transfer to Mixset and Katoh--Suzuki---use the main fits trained on the
2182-guide Huesken training set. In the fourth, all fitted models are re-trained on 489 guides
against rodent inserts and evaluated on 1453 guides against human inserts; on this training set,
Silscore retains $\mathrm{U1}$ and $\mathrm{G18}$. Section~\ref{sec:targets} examines this unseen-target evaluation
separately.
The upper row of Figure~\ref{fig:performance} reports the correlation for each method and its
separate 95\% percentile-bootstrap interval, calculated by resampling the evaluation records
with the fitted models held fixed. For OligoFormer, the bootstrap also resamples the training--validation
splits and PyTorch seeds before averaging the corresponding run correlations, so that the intervals account for both the
limited test set and variation between training runs (Supplementary Section~\ref{sec:comparators}).
These intervals include variation in which evaluation records are resampled, and the same easy
or difficult records tend to move all methods in the same direction. Their overlap therefore
does not quantify the difference between two methods; rather, it reflects uncertainty in each
method's performance as the composition and difficulty of the evaluation sample vary. The lower
row estimates that difference directly: in each of 4000 bootstrap replicates, the same evaluation
records are resampled for both methods before their correlations are subtracted.

Silscore has the highest observed correlation in all three transfer evaluations. On Katoh--Suzuki,
it ranks above every applicable comparator, including the best DSIR representation available
there by $+0.023$ $[+0.0002, +0.045]$. On Mixset and unseen targets, however, the data do not
establish that it outperforms the closest comparators. The held-out Huesken split is the only
evaluation where Silscore does not have the highest observed correlation, but the data likewise do not establish
that the fitted comparators outperform Silscore. As discussed above, however, this within-Huesken comparison may give an optimistic
estimate because it neither tests robustness to changes in experimental context nor controls
train--test sequence similarity. Across all four
evaluations, the bootstrap comparisons support an advantage for Silscore over both rule-based scores.

The comparison of model variants also quantifies the contribution of the selected
identity terms beyond the Thermo-only model: Silscore outperforms Thermo-only in all four
evaluations, with gains of $+0.064$ on held-out Huesken, $+0.064$ on Mixset, $+0.032$ on
Katoh--Suzuki and $+0.068$ on unseen targets; the bootstrap comparisons support an improvement in
every evaluation. The $\mathrm{U1}$
preference accounts for most of this contribution on held-out Huesken,
where the five additional terms add only $+0.004$ $[-0.026, +0.035]$ over Thermo+$\mathrm{U1}$.
Notably, however, these terms improve transfer by $+0.037$ $[+0.014, +0.060]$ on Mixset and
$+0.026$ $[+0.007, +0.045]$ on Katoh--Suzuki, supporting their transferability beyond the
training data.

\subsection{Silscore retains performance under reduced training--test target similarity}\label{sec:targets}

\begin{figure}[!htbp]
\centering
\includegraphics[width=0.8\textwidth]{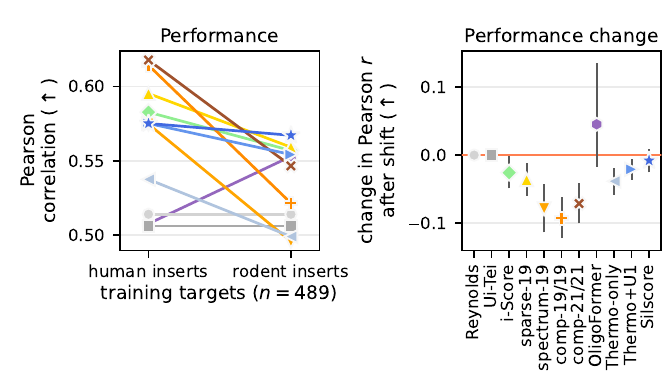}
\caption{Performance on the same 1453 siRNAs against human reporter inserts after training on
guides against human or rodent inserts; all measurements come from the same human H1299 assay.
Left: Pearson correlations in the two training conditions; Reynolds and Ui-Tei are unchanged
because they are fixed rules and are not fitted. Right: the change in Pearson
correlation when training changes from human to rodent inserts (rodent-trained minus human-trained),
with positive values indicating a gain and negative values a loss. Intervals are 95\% bootstrap
intervals over the common evaluation records. The
bootstrap construction is the same as in Figure~\ref{fig:performance}, but each method is
compared with itself rather than with Silscore. OligoFormer shows mean correlations across 25
runs; the same split and seed indices are resampled in both training conditions. Colours identify the same methods as in Figure~\ref{fig:performance}; marker
shapes identify methods consistently across the two panels.}
\label{fig:species}
\end{figure}

Comparing performance on the Huesken test set with performance on the external datasets does
not isolate the effect of training--test target similarity, because the evaluation dataset---and
with it the assay conditions and sequence composition---also changes. We
therefore evaluated all models on the same 1453 siRNAs targeting human reporter
inserts---target-derived sequences placed in the reporter transcripts---after training them on either
the remaining 489 siRNAs targeting human inserts or the
489 targeting rodent inserts. Replacing the human-insert training set with the
rodent-insert set changes the sequence information available to every fitted model. The human-insert training and evaluation sets are sampled from the same
collection of reporter inserts. At the level of nucleotide composition, their mean guide GC contents
are $49.7\%$ and $49.4\%$, respectively, whereas the rodent-insert training set has a mean guide GC
content of $52.8\%$. At the level of local sequence overlap, $39\%$ of the evaluation guides
contain a contiguous stretch of at least 13 nucleotides already found in the human-insert training set. Against the rodent-insert training set, the same is true
for only 1 of 1453 evaluation guides ($0.07\%$).
This provides a composite stress test in
which both sequence composition and local sequence familiarity change while the evaluation
records remain fixed. All
measurements were made in human H1299
cells, so human and rodent refer only to the origin of the reporter insert, not to the
cellular system; assay, laboratory, model class and selection procedure were also held fixed.
The identity-term selection procedure was re-applied separately to each of the two 489-guide training sets.
It retained $\mathrm{U1}$ alone on the human-insert guides and $\mathrm{U1}$ plus $\mathrm{G18}$ on the rodent-insert
guides, giving Silscore 21 and 22 fitted parameters, respectively.
Figure~\ref{fig:species} reports performance under both training conditions and the paired
within-model change, using a bootstrap analogous to the construction of Figure~\ref{fig:performance}.

Silscore's correlation changes by $-0.008$ $[-0.025, +0.009]$. The data do not establish a loss for
Silscore, whereas they support a loss for DSIR, i-Score and the two thermodynamic variants.
Unexpectedly, OligoFormer is the only fitted model with an observed increase: its mean correlation
rises from $0.508$ after human-insert training to $0.553$ after rodent-insert training. However,
the interval for the change, $+0.045$ $[-0.017, +0.135]$, includes zero, so the data do not
establish an improvement and we do not interpret its direction.
All DSIR variants show larger
observed decreases, with sparse-19 the closest, a
pattern consistent with Silscore's smaller number of fitted parameters.
After the shift, Silscore also has the highest observed Pearson correlation among the fitted
methods, although the paired bootstrap comparisons do not establish a clear advantage over OligoFormer, sparse-19 or i-Score.

The three Silscore variants link part of this robustness to the identity terms: the observed change approaches zero,
from $-0.038$ for Thermo-only to $-0.021$ for Thermo+$\mathrm{U1}$ and $-0.008$ for Silscore. Thus $\mathrm{U1}$
provides part of the gain, while the complete selection protocol improves robustness further.
This comparison therefore isolates the component of transfer associated with changing the
training targets from human to rodent inserts, while holding assay, laboratory and evaluation set fixed. Cross-dataset transfer
changes these factors jointly. Because they change together, the available data cannot show how much
each factor contributes to the change in performance or whether different models are affected to the
same extent.

\subsection{Overall duplex stability has an intermediate optimum}
\label{sec:global-stability}

The quadratic term in Equation~\ref{eq:score} allows the model to learn an optimal duplex
stability $\dG^*$ and a penalty strength $\kappa$. We impose no prior constraint on either parameter,
so both are estimated from the
data rather than inherited from previous design rules. We applied the complete fitting procedure
separately to the 2182 siRNAs in the Huesken training set and to all labelled siRNAs in Mixset
and Katoh--Suzuki. The 249 Huesken test siRNAs remain excluded. These separate fits are used only to describe the stability relationship
within each dataset; they are not used for the predictive comparisons below.
Because the efficacy scales differ across assays, we divide each fitted value of $\kappa$ by the
standard deviation of efficacy within that dataset before comparing datasets.

\begin{figure}[!htbp]
\centering
\includegraphics[width=0.90\textwidth]{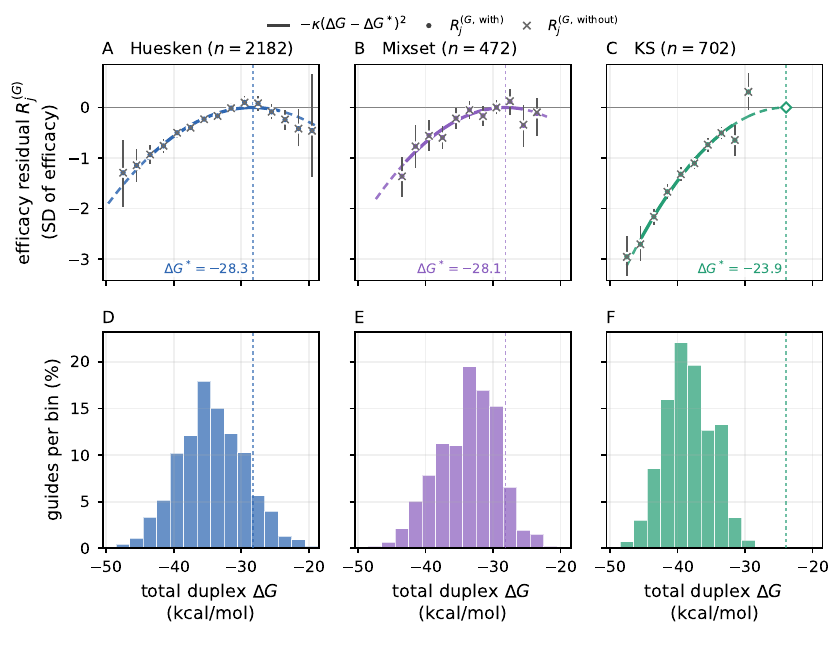}
\caption{Fitted quadratic stability term and the residuals it explains in Huesken (A, D), Mixset
(B, E) and Katoh--Suzuki (C, F).
The top row compares the fitted penalty $-\kappa(\dG-\dG^*)^2$ with binned means of the
residual $R_j^{(G)}$ defined in Equation~\ref{eq:global-partial-response}. Both quantities are
divided by the standard deviation of efficacy within each dataset. Circles denote $R_j^{(G,\,\mathrm{with})}$,
computed from the fit with the quadratic term; crosses denote
$R_j^{(G,\,\mathrm{without})}$, computed after refitting without the quadratic term while
retaining the term linear in $\dG$. The latter are shifted by a single additive constant so that the
two sets share the vertical reference of the full fit; the shift leaves their shape unchanged.
The same
2\,kcal/mol bins are used in both rows; bins with fewer than five guides are
omitted from the top row. Error bars are 95\% within-bin bootstrap intervals for the circles,
with the fit held fixed. The bottom row shows the percentage of guides in
each bin. Curves are solid over
the central 95\% of the observed $\dG$
distribution and dashed in the observed tails or beyond the data. Vertical dashed lines mark
the fitted optima; in C the open diamond marks the extrapolated optimum. Each dataset uses a
separate fit to the guides shown in its histogram.}
\label{fig:global-stability}
\end{figure}

Huesken and Mixset gave compatible, nearly identical estimates of the optimum:
$-28.3$ [$-29.7$, $-26.0$] and $-28.1$ [$-30.6$, $-21.5$]\,kcal/mol, respectively
(\mbox{Figure~\ref{fig:global-stability}A--B}). The Mixset interval is substantially wider, so its
estimate provides weaker localisation of the optimum. Both estimates fall within the empirical
$-35$ to $-27$\,kcal/mol range that \citet{matveevaOptimizationDuplexStability2010} associated
with the highest mean efficacy across four independent siRNA datasets. That study used the same
nearest-neighbour parameter set to calculate $\dG$. Both fitted values of $\kappa$ were positive,
with 95\% bootstrap intervals above zero and overlapping standardised estimates:
$0.0041$ [$0.0030$, $0.0053$] and $0.0049$ [$0.0023$, $0.0076$]
$(\mathrm{kcal/mol})^{-2}$ for Huesken and Mixset, respectively. Both datasets
contain observations on either side of the fitted optimum, although the less-stable side
($\dG>\dG^*$) comprises only 256 of 2182 Huesken guides (12\%) and 43 of 472 Mixset guides
(9\%).

The Katoh--Suzuki data are compatible with an intermediate optimum but cannot locate it, because
they sample only the more-stable side of the relationship: the least stable guide has
$\dG=-29.5$\,kcal/mol, still more stable than either the Huesken or Mixset estimate. Its
fitted optimum therefore lies outside the data at $-23.9$\,kcal/mol, with a wide bootstrap interval
[$-29.9$, $+4.6$]\,kcal/mol; only $3.7\%$ of bootstrap estimates fall within the observed range
(Figure~\ref{fig:global-stability}C). We therefore do not interpret this estimate as a
Katoh--Suzuki-specific optimum. The fitted value of $\kappa$ is positive and compatible with the other datasets,
$0.0056$ [$0.0018$, $0.0094$] $(\mathrm{kcal/mol})^{-2}$ after the same standardisation,
but because Katoh--Suzuki covers only the more-stable side of the relationship, the observed
trend could also be explained by a more general monotonic dependence.

To assess the predictive contribution of overall duplex stability---that is, the linear and
quadratic terms in $\dG$---we return to the main
evaluation setup: all model variants are fitted on Huesken alone, then evaluated on the held-out
Huesken test set and transferred without re-fitting to Mixset and Katoh--Suzuki. Removing both
the linear and quadratic terms in $\dG$ reduced Pearson correlation by $0.048$
[$0.010$, $0.090$] on the Huesken test set, $0.035$ [$0.005$, $0.065$] on Mixset and $0.183$
[$0.149$, $0.216$] on Katoh--Suzuki. Relative to a model retaining only the term linear in $\dG$, the
quadratic component learned on the Huesken training records improved correlation by $0.021$ [$-0.001$, $0.044$], $0.014$
[$0.004$, $0.025$] and $0.065$ [$0.051$, $0.079$], respectively. The overall duplex-stability terms
therefore improve prediction on held-out Huesken and on both transfer datasets. The data support
a contribution from the quadratic term on both external datasets. On held-out Huesken, its estimated contribution
is positive, but the data remain compatible with no improvement. In these comparisons, the
six selected identity terms are held fixed. A control that removes the overall duplex-stability terms and
repeats identity selection is reported in Supplementary Section~\ref{sec:global-stability-sensitivity}.

Using these three separate fits, which are used only to describe the stability relationship within
each dataset, we next ask what pattern the quadratic stability term captures. Subtracting the fitted intercept,
local-stability profile and selected identity contributions
from each observed efficacy leaves the residual to be explained by $-\kappa(\dG-\dG^*)^2$:
\begin{equation}
\label{eq:global-partial-response}
R_j^{(G)} \equiv y_j-
\left[
\beta_0 + \sum_{i=1}^{18} w_i \dG_{j,i}
+ \sum_{p=1}^{19}\theta_p(g_{j,p})
\right] .
\end{equation}
In Figure~\ref{fig:global-stability}, we group guides into 2\,kcal/mol intervals of $\dG$, average
their residuals within each interval, and compare these averages with the fitted penalty
$-\kappa(\dG-\dG^*)^2$. The curve closely tracks the mean residual dependence
on $\dG$: correlations with the binned means range from $0.94$ to $0.98$ across the three
datasets. To test whether the quadratic term itself produces this pattern, we refitted the model
without it, retaining the term linear in $\dG$ together with the local-stability and identity
terms. The resulting binned residual profile was nearly unchanged and was still well
described by the quadratic curve; these residuals are overlaid as crosses in
Figure~\ref{fig:global-stability}. Their correlations with
the quadratic curve differ from the full-fit values by less than $0.003$. The same holds at the
level of individual guides, before averaging within intervals: the correlations change by at
most $0.0014$ when the quadratic is removed. Averaging within intervals therefore exposes a mean dependence on total duplex stability that the quadratic describes
well. Because essentially the same residual pattern is obtained from a fit that does not contain
the quadratic term, the agreement is not a circular artefact of the joint fit: the quadratic
tracks an existing $\dG$-dependent structure rather than producing it by reallocating the other
components. At the level of individual guides, without averaging within intervals, the corresponding correlations
are $0.329$, $0.339$ and
$0.613$ in Huesken, Mixset and Katoh--Suzuki. The lower values reflect variation among guides with similar
duplex stability that is averaged out within each interval. This remaining variation may include
effects not represented by the model as well as experimental noise; the mean dependence on
$\dG$ nevertheless emerges clearly once this within-interval variation is averaged.

Together, the predictive gains from including overall duplex stability in the held-out and
transfer evaluations, and the compatible Huesken and Mixset optima, support a dependence on
overall duplex stability. Adding a cubic coefficient while holding the six identity terms fixed
changed Pearson correlation by $+0.002$ [$-0.005$, $+0.009$] on the Huesken test set, but by
$-0.003$ [$-0.008$, $+0.001$] on Mixset and $-0.009$ [$-0.014$, $-0.003$] on Katoh--Suzuki. Thus the
small gain on held-out Huesken did not transfer to the external datasets. We therefore retained the
quadratic as an effective empirical summary of the observed relationship, without assigning
mechanistic significance to its exact form. It
turns a previously reported preference for intermediate stability into a smooth penalty whose
optimum, strength and uncertainty are estimated from the data rather than prescribed as thresholds.
\subsection{The fitted energy profile recovers regional preferences}
\label{sec:weight-profiles}

Predictive performance does not by itself show whether the fitted physical parameters are
interpretable. We therefore inspect the nucleotide profile $W_i$ defined in
Equation~\ref{eq:nucleotide-weight}. We applied the complete procedure independently to Huesken, Mixset and
Katoh--Suzuki. The displayed profiles come from the complete fits, including the identity terms selected within each dataset.
To separate physical and identity contributions, we compare them below with Thermo-only profiles fitted without identity terms.
These dataset-specific fits are used only to compare the fitted profiles; they are not part of the
transfer evaluation above. Because the three assays do not share an absolute
activity scale, each profile is expressed per standard deviation of the efficacy measured in
that dataset. After accounting for overall duplex stability, a positive $W_i$ favours weaker
local pairing at that position, whereas a negative $W_i$ favours stronger local pairing. The
underlying step weights $w_i$ sum to zero, so the profile describes how local-stability effects
are distributed among positions rather than the absolute effect of each position.

\begin{figure}[!htbp]
\centering
\includegraphics[width=0.92\textwidth]{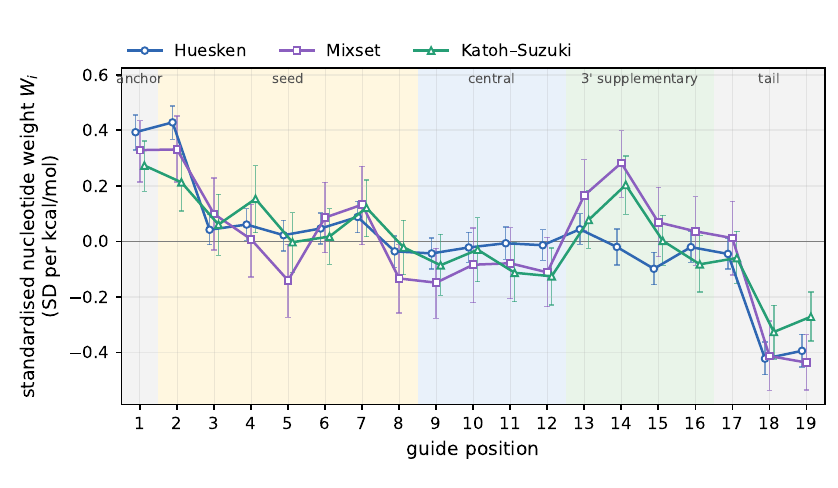}
\caption{Silscore local-stability profiles fitted independently to Huesken, Mixset and Katoh--Suzuki.
$W_i$ is the sum of the weights of the dinucleotide steps that contain guide nucleotide $i$,
divided by the within-dataset standard deviation of efficacy. Positive values favour weaker
local pairing relative to the global-stability term. Error bars are pointwise 95\% intervals
from 4000 bootstrap samples of the records used for each fit: the 2182 Huesken training records
and all labelled Mixset and Katoh--Suzuki records. Intervals condition on the identity terms selected
from the original sample. Background colours mark the Argonaute guide regions defined by
\citet{weeArgonauteDividesIts2012} and discussed in the text. The selection retains six identity terms in Huesken and
none in Mixset or Katoh--Suzuki.}
\label{fig:weight-profiles}
\end{figure}

All three profiles recover the previously reported terminal asymmetry: their fitted weights favour weaker
pairing at the guide $5'$ end and stronger pairing at its $3'$ end
(Figure~\ref{fig:weight-profiles}). This
is the largest feature of every profile and agrees with thermodynamic asymmetry and sequence
rules reported previously \citep{ui-teiGuidelinesSelectionHighly2004,
reynoldsRationalSiRNADesign2004,vertAccurateInterpretableModel2006}. Here it emerges as a
continuous energy profile, without encoding terminal base preferences as physical rules.

A smaller regional structure recurs between the ends. Using the definitions of the Argonaute guide domains from
\citet{weeArgonauteDividesIts2012}, the mean Silscore weight is higher in the seed (positions 2--8)
than in the central region (positions 9--12) by $0.115$, $0.160$ and $0.166$ standard deviations per
kcal/mol in Huesken, Mixset and Katoh--Suzuki, respectively. Thus, weaker local pairing is
predicted to contribute more positively to potency in the seed than in the central region,
as quantified by these differences in the fitted weights. This preference is compatible with the
earlier preference for an A/U-rich guide $5'$ region
\citep{ui-teiGuidelinesSelectionHighly2004} and with the weak-stability regions identified by
\citet{vertAccurateInterpretableModel2006}. The local shape is not identical, but maxima recur within
the seed, including near position 7 in each dataset. These repeated peaks are suggestive, but
their uncertainty is too large to justify new design rules for individual positions. We therefore
interpret them as part of the broader difference between the seed and central regions.

Agreement across independently fitted datasets is high for the complete profiles: Pearson
correlations are $0.85$ between Huesken and Mixset, $0.86$ between Huesken and
Katoh--Suzuki, and $0.92$ between Mixset and Katoh--Suzuki. The most conspicuous local
difference is at position 14, where Huesken lies near zero while Mixset and Katoh--Suzuki are positive.
Huesken is also the only fit in which $\mathrm{G14}$ passes the identity-term threshold. Thus the
difference can arise from how correlated physical and identity contributions are allocated by
the joint fit. Indeed, the Huesken Thermo-only and Thermo+$\mathrm{U1}$ weights at position 14 are
positive and closer to Mixset and Katoh--Suzuki (Supplementary
Figure~\ref{fig:weight-profiles-supp}B--D). The combined physical and identity contributions at
this position are shown in the nucleotide map of Section~\ref{sec:nucleotide-preferences}.

The same broad profile is present before adding identity terms (Supplementary
Figure~\ref{fig:weight-profiles-supp}A--D). In Huesken, the Thermo-only and Silscore profiles
correlate at $0.98$, whether or not position 1 is included. Their largest reallocation is at
position 1, where Silscore can assign the known nucleotide-specific contribution to the
identity terms at that position, dominated by $\mathrm{U1}$. On Mixset and Katoh--Suzuki no identity term is selected,
so Silscore and Thermo-only coincide. We also report the raw
dinucleotide-step weights from which the $W_i$ are derived (Supplementary
Figure~\ref{fig:weight-stacks-supp}). Taken together, these results show that the fitted profile
recovers known preferences and extends them across datasets in a common, data-driven
thermodynamic representation.

\subsection{Identity terms selected from data}\label{sec:identity}

We apply the criterion of Section~\ref{sec:identity-selection} to select the identity terms.
The selection and coefficient estimates reported here use only the 2182 Huesken training siRNAs;
the held-out 249, Mixset and Katoh--Suzuki play no part in the selection, keeping the performance
comparisons in Section~\ref{sec:performance} independent of it. All 76 position--base combinations are tested (19 positions $\times$ 4 bases,
with even the least frequent combination occurring in 468 of the 2182 guides), and six pass the
multiplicity-corrected threshold defined in the Methods and are reported in
Table~\ref{tab:identity}. For each combination, $z$ measures the standardised departure of the
mean Thermo-only residual from zero. After selection, $\theta$ is the fitted coefficient obtained
when the identity terms and physical parameters are refitted jointly. Thus $z$ identifies a
deviation unexplained by the thermodynamic model, whereas $\theta$ quantifies its contribution to
the final score; neither provides an absolute ranking of the four bases at a position.

\begin{table}[!htbp]
\centering
\begin{tabular}{lrrr}
\toprule
identity term & $z$ & $\theta$ & guides \\
\midrule
$\mathrm{U}$ at 1  & $+9.4$ & $+0.092$ & 621 \\
$\mathrm{A}$ at 19 & $-6.1$ & $-0.061$ & 495 \\
$\mathrm{G}$ at 18 & $-5.2$ & $-0.053$ & 581 \\
$\mathrm{C}$ at 2  & $+4.0$ & $+0.027$ & 521 \\
$\mathrm{C}$ at 1  & $-3.6$ & $-0.019$ & 510 \\
$\mathrm{G}$ at 14 & $-3.5$ & $-0.053$ & 559 \\
\bottomrule
\end{tabular}
\caption{The six position--base identity terms selected on the Huesken training set, ordered by
$|z|$. Here $z$ is the screening statistic for the mean Thermo-only residual, and $\theta$ is the
coefficient after joint refitting, in units of activity. The last column gives the number of the
2182 training guides carrying that base at that position. Because the model is refitted after
selection, $\theta$ need not follow $z$ in magnitude or sign.}
\label{tab:identity}
\end{table}

The strongest selected identity term is $\mathrm{U}$ at position 1, and it corresponds to a known
mechanistic determinant: Argonaute reads the guide $5'$ nucleotide through a protein contact
\citep{frankStructuralBasis5nucleotide2010}. Because this interaction is not a base-pairing
effect, it cannot be captured by the stacking energies, and the model represents it separately
through an identity term. The model also selects $\mathrm{C1}$ with the opposite sign. Because the four
bases at a position are mutually exclusive and their weighted residual means sum to zero, a
strong $\mathrm{U1}$ signal necessarily produces compensating deviations among the non-$\mathrm{U}$ bases. We
therefore treat $\mathrm{U1}$ and $\mathrm{C1}$ as one position-1 signal, dominated by $\mathrm{U1}$ ($z=+9.4$), rather than
treating $\mathrm{C1}$ as an independent mechanistic finding. The strongest selected deviation outside
position 1 is $\mathrm{A}$ at position 19. Guides carrying $\mathrm{A}$ or $\mathrm{U}$ at this position have nearly the
same mean duplex free energy ($-32.99$ and $-33.36$ kcal/mol, respectively), yet differ in
activity in a way that the thermodynamic component does not capture.

All six terms fall either at the ends of the guide (positions 1--2 and 18--19) or at
position 14, within the supplementary-pairing region spanning positions 13--16
\citep{weeArgonauteDividesIts2012}. These locations are therefore biologically plausible.
These terms were selected from the data rather than proposed in advance, and assigning a mechanism
to each would require independent validation beyond the scope of this paper. Some may compensate
for limitations of the thermodynamic representation rather than reflect independent effects of
nucleobase identity, and some may be false positives despite the multiplicity correction.

\subsection{Variance across the three model terms}\label{sec:complete-model}
We can interrogate the model fitted on the 2182 Huesken training guides to assess the
relative role of each term in the score. For this, we examine how much each term varies
across guides: a higher variance means larger differences in that term's contribution
to their scores. We apply the same fitted model, without refitting, to the Huesken test
set, Mixset and Katoh--Suzuki. Supplementary Figure~\ref{fig:term-covariance} shows
the variances and covariances of the three terms.

Across all three datasets, local stability has the largest variance, roughly two to three
times that of either global stability or nucleotide identity, suggesting a substantial
contribution to differences in predicted potency between guides.
Despite having only six selected coefficients, their combined contribution has a
variance comparable to that of global stability.
The covariances are small relative to the component variances: global stability covaries weakly
and positively with the other two components, whereas local stability and nucleotide identity
are essentially uncorrelated.

\subsection{Summary of learned preferences}\label{sec:nucleotide-preferences}
\begin{figure}[!htbp]
\centering
\includegraphics[width=0.86\textwidth]{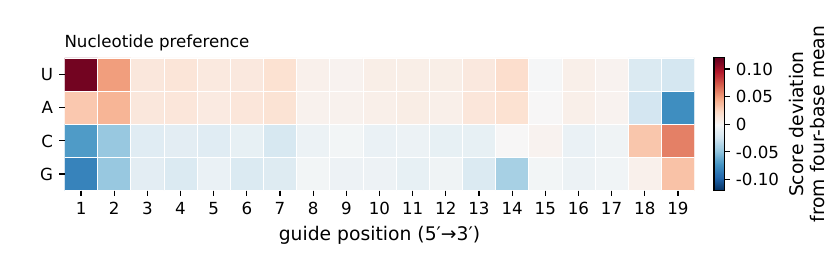}
\caption{Nucleotide preferences of the complete Huesken-trained model. Each cell shows the
mean predicted score after setting the indicated base at that position in all 249 Huesken
test guides, minus the mean across the four bases at the same position. The other 18 guide
nucleotides and all fitted coefficients are held fixed; each variant is evaluated as a
perfectly paired duplex. Red and blue indicate scores above and below the four-base mean.}
\label{fig:nucleotide-heatmap}
\end{figure}

We have examined the model terms individually, but their nucleotide preferences can be
misleading when interpreted in isolation. We therefore examine how the terms of the
Huesken-trained model together shape nucleotide preferences in the total score.
We use all 249 guides in the Huesken test set as starting sequences. At each position,
we replace the nucleotide in every guide with each of $\mathrm{U}$, $\mathrm{A}$, $\mathrm{C}$ and $\mathrm{G}$, keeping the
other 18 nucleotides unchanged, and average the model's predicted scores over the 249
sequences for each replacement base. Each cell in Figure~\ref{fig:nucleotide-heatmap}
shows this average minus the mean of the four base-specific averages at that position.
A positive value therefore means that the model favours that base relative to the
four-base average, across the test-set sequences.

The map recovers
terminal asymmetry, with $\mathrm{A/U}$ favoured across much of the $5'$ region and $\mathrm{C/G}$ at the
$3'$ end, together with a strong preference for $\mathrm{U1}$ and a strongly negative value for $\mathrm{G14}$.
The $\mathrm{A/U}$ preference across much of the seed (positions 2--8) also reflects the regional
preference for weaker seed pairing discussed in Section~\ref{sec:weight-profiles}.
The $\mathrm{C2}$ term illustrates why the combined view matters: despite its positive identity
coefficient, $\mathrm{C}$ at position 2 remains below the four-base mean in the complete score.

To assess whether these patterns depend on the particular sequences used, we repeat the
same substitutions and averaging over the Mixset and Katoh--Suzuki guides, keeping the
Huesken-trained model fixed. The resulting maps are qualitatively similar across these
different sets of guide sequences (Supplementary Figure~\ref{fig:nucleotide-heatmap-supp}).

In practice, the target transcript is usually specified, and the available binding sites
constrain the candidate guide sequences. These maps help interpret what the model has
learned rather than provide design rules for choosing bases independently. Moreover, their entries
are not additive: stacking interactions and the global-stability penalty couple positions,
so each candidate must be evaluated with the complete score.

%% file: sections/discussion.tex
\section{Discussion}\label{sec:discussion}

Here we present Silscore, a compact model for siRNA potency prediction that combines positional thermodynamics with
sparse nucleotide-identity terms selected from data. Across the two external datasets, Mixset and
Katoh--Suzuki, and the unseen-target evaluation within the Huesken collection, Silscore has the highest
observed Pearson correlation among the methods compared. On Katoh--Suzuki, the data support an
advantage over every applicable comparator. On Mixset and unseen targets, the comparisons with
the closest methods are inconclusive, although Silscore ranks first in both. The original Huesken
train--test split is the only evaluation in which more flexible models have modestly higher
observed correlations, but the bootstrap comparison does not show a clear difference from
Silscore for any of them. Because the original Huesken split assigns individual siRNAs, rather than whole targets,
to training or evaluation, it retains the greatest overlap in experimental design and is the
comparison most likely to reward additional flexibility.

Silscore's main fit uses 26 parameters, fewer than half as many as any trainable comparator
evaluated here.
When the same human-insert guides were used for evaluation and the training set
was changed from human inserts in the same collection to distinct rodent inserts, the data did not
establish a performance change for Silscore, which had the highest observed correlation
after the shift.

The global-stability analysis identifies a consistent intermediate optimum. Independent fits to
Huesken and Mixset place it at nearly the same value, around $-28.2$ kcal/mol, within the
intermediate range previously associated with high siRNA efficacy
\citep{matveevaOptimizationDuplexStability2010}.
Including both overall duplex-stability terms improves predictions on the held-out Huesken data
and on both external datasets. The dependence on total duplex stability is already visible
in the residuals of models fitted without the quadratic term, and the quadratic penalty describes
it closely after the residuals are averaged within duplex-energy intervals.
The quadratic is not intended as a mechanistic model. It is a simple smooth function that
represents a dependence with an intermediate optimum, fits the observed pattern closely, and
provides estimates of the optimum and penalty strength.

Positional profiles of stacking energies
were explored previously but were less accurate on their own than the sequence representations
compared with them \citep{vertAccurateInterpretableModel2006}. Silscore instead uses the profile as
one component of a broader model. Although the components are fitted jointly within each model,
fits performed independently on each dataset recover the previously reported terminal asymmetry
\citep{khvorovaFunctionalSiRNAsMiRNAs2003,schwarzAsymmetryAssemblyRNAi2003,
reynoldsRationalSiRNADesign2004}. They also consistently favour weaker pairing in the seed than
in the central region, in agreement with earlier preferences for an
$\mathrm{A/U}$-rich guide $5'$ region
\citep{ui-teiGuidelinesSelectionHighly2004,vertAccurateInterpretableModel2006}. The Silscore
profiles are highly correlated across the three datasets, despite differences in their targets and
measurements.
Smaller position-to-position variations are less consistent and should not be overinterpreted: the
present statistics do not support turning individual peaks into new position-specific rules. The
clearer result is that known empirical preferences, together with a broader difference between
regions, emerge from independently fitted energy profiles rather than being encoded as scoring rules.

The data-selected identity terms also make an important contribution to prediction. Only six
additional coefficients are retained in the main fit, yet adding them improves on Thermo-only in
all four evaluations, and the bootstrap comparisons support an improvement in every case.
Silscore selects these terms from residuals on the Huesken training set alone; beyond the known
$\mathrm{U1}$ preference, the remaining five improve predictions on both external datasets, with the
bootstrap analysis supporting an improvement on both. The strongest term, a preference for $\mathrm{U}$ at
the guide $5'$ nucleotide, is consistent with direct nucleotide-specific recognition by Argonaute
\citep{frankStructuralBasis5nucleotide2010}. The remaining terms concentrate near the guide ends
and at position 14, within the $3'$ supplementary region spanning positions 13--16
\citep{weeArgonauteDividesIts2012}. These locations are biologically plausible and provide
specific hypotheses for further study. Assigning a distinct mechanism to each coefficient is
beyond the present scope, but their improvement on datasets that played no part in their selection
shows that they contain predictive information beyond the thermodynamic component and the $\mathrm{U1}$
term alone.

The explicit score components also make their relative magnitudes directly accessible.
In the Huesken-trained model, local stability has the largest variance
across the three evaluation datasets (Supplementary Figure~\ref{fig:term-covariance}), while the
combined contribution of the six identity terms has a variance comparable to that of global stability. A sparse identity correction can therefore
make a substantial contribution to the differences in predicted score between guides.
The nucleotide map shows how these contributions combine into familiar sequence preferences,
including the terminal asymmetry, while resolving distinctions such as the strong $\mathrm{U1}$
preference. Together with the held-out ablations, these analyses connect predictive utility,
the magnitude of the fitted contributions and their sequence dependence within one model.

Silscore is deliberately guide-only and therefore estimates the intrinsic sequence contribution
to potency. Transcript-level effects such as target-site accessibility are not represented. The
model was also fitted and evaluated on unmodified siRNAs, using nearest-neighbour parameters
defined for unmodified RNA duplexes. Because Silscore is physics-based, its fitted relationships
may remain informative for chemically modified siRNAs, and the framework could also
incorporate modification-specific thermodynamic parameters. Neither possibility is tested here;
both would require dedicated validation, because chemical modifications can affect duplex
energetics and other determinants of potency \citep{bofill-derosLearningRulesThat2023,
davisSystematicAnalysisSiRNA2025}.

Silscore therefore provides a compact, data-driven model that combines strong predictive
performance with robustness when tested in new contexts. Its parameters can also be interpreted
directly: the global term quantifies both the optimal stability and the decrease in predicted
potency away from it, the local profile estimates how pairing stability should be distributed along
the guide, and the sparse identity terms locate sequence effects that cannot be
expressed through the base-pairing energies used here. Together, these components express several
previously reported empirical preferences as quantities estimated from data within a single model.

%% file: sections/methods.tex
\section{Methods}\label{sec:methods}

\subsection{Thermodynamic calculations}\label{sec:thermodynamic-calculations}

Guides were represented as 19-nt RNA sequences in the antisense $5'$-to-$3'$ orientation.
We computed $\Delta G^\circ_{37}$ using the Watson--Crick RNA:RNA nearest-neighbour parameter set of \citet{xiaThermodynamicParametersExpanded1998}, as tabulated in the Nearest Neighbor Database \citep{turnerNNDBNearestNeighbor2010}.
For a 19-nt duplex, the overall free energy is
\begin{equation}\label{eq:duplexdg}
\dG(g) = \dG_{\mathrm{init}} + n_{\mathrm{AU}}\,\dG_{\mathrm{AU}}
+ \sum_{i=1}^{18} \dG_i ,
\end{equation}
where $\dG_{\mathrm{init}}=4.09$ kcal/mol is the initiation term,
$n_{\mathrm{AU}}\in\{0,1,2\}$ is the number of terminal $\mathrm{A}\!\cdot\!\mathrm{U}$ pairs, and
$\dG_{\mathrm{AU}}=0.45$ kcal/mol is the penalty for each such pair.
The guide was paired to its exact 19-nt complement; overhangs, dangling ends, mismatches and secondary structures were not included.

\subsection{Model fitting}\label{sec:model-fitting}

For a selected set $S$ of identity terms, we estimate the model of
Equation~\ref{eq:score} by minimising
\begin{equation}
\sum_j\bigl(y_j-\hat y(g_j)\bigr)^2
+ \lambda\sum_{i=1}^{18}w_i^2
+ \mu\sum_{(p,b)\in S}\theta_{p,b}^2,
\qquad \text{subject to } \sum_{i=1}^{18}w_i=0 .
\end{equation}
The intercept and the global-stability term are not penalised.
The constrained ridge problem is solved in closed form. Thermo-only, Thermo+$\mathrm{U1}$ and Silscore
have 20, 21 and, for the main Huesken fit, 26 free parameters, respectively.
In the command-line implementation, fitting the final model on the 2182 Huesken training guides
and scoring the 82 candidates from a 100-nt target each took under one second on an Apple M1 CPU.

For the nucleotide-level profiles, we combine the fitted weights of the dinucleotide steps
adjoining each nucleotide. For internal positions,
\begin{equation}\label{eq:nucleotide-weight}
W_i=w_{i-1}+w_i, \qquad i=2,\ldots,18,
\end{equation}
with $W_1=w_1$ and $W_{19}=w_{18}$.

The two hyperparameters $\lambda$ and $\mu$ were chosen by evaluating a logarithmic grid
whose consecutive positive values differed by a factor of $\sqrt{10}$, from $10^{-3}$ to
$10^5$ for $\lambda$ and from $10^{-3}$ to $10^4$ for $\mu$, together
with the unpenalised $\lambda=0$ and $\mu=0$ axes, repeating
five-fold cross-validation ten times on the 2182 Huesken training records. Identity-term
selection was repeated within every training fold. The surface contained a broad plateau,
with the largest mean pooled out-of-fold Pearson correlation at $(\lambda,\mu)=(1,0)$. At $(1,1)$,
the mean correlation was lower by only $1.9\times10^{-5}$. Given this practically equivalent
performance, we used $\lambda=1$ and $\mu=1$, preferring the more regularised solution that
penalises both coefficient blocks. This choice lies within the plateau and away from the region
where performance declines
(Supplementary Figure~\ref{fig:hparam-landscape}).

\subsection{Selection of identity terms}\label{sec:identity-selection}

Identity terms are selected using the training data only. We first fit Thermo-only, the model without
identity terms, and compute its residuals. This model includes the free unpenalised intercept
$\beta_0$ of Equation~\ref{eq:score}, which makes their mean zero to machine precision.
Because the four bases partition the guides at each position, their weighted residual means
also sum to zero. Multiple retained cells at one position are therefore components of a joint
contrast, not independent findings.
For each position $p\in\{1,\ldots,19\}$ and base
$b\in\{\mathrm{A},\mathrm{C},\mathrm{G},\mathrm{U}\}$, we collect the residuals of guides for which $g_p=b$ and compute
\begin{equation}
z_{p,b}=\frac{\bar{e}_{p,b}}{s_{p,b}/\sqrt{n_{p,b}}},
\end{equation}
where $\bar{e}_{p,b}$, $s_{p,b}$ and $n_{p,b}$ are their mean, sample standard deviation and
number of guides. Combinations represented by fewer than 30 guides are not tested. If $m$ combinations are
tested, we retain those satisfying the two-sided Bonferroni criterion \citep{dunnMultipleComparisonsMeans1961}
\begin{equation}
|z_{p,b}| \geq \Phi^{-1}\!\left(1-\frac{0.05}{2m}\right).
\end{equation}
All 76 combinations exceed the minimum count in the main Huesken training set, giving a threshold of
3.41.
After selection, the retained identity coefficients and all physical parameters are refitted
jointly; the remaining identity coefficients stay fixed at zero. Thus $z$ is a screening
statistic, whereas $\theta$ is a coefficient of the final joint fit, and their ranking or sign
need not agree. The complete selection and fitting procedure is repeated whenever the training
set changes.

Unless otherwise stated, parameter and profile intervals use 4000 nonparametric bootstrap
resamples of the records fitted, conditional on the identity terms selected from the original
sample. Performance differences use paired resamples of the common evaluation records with the
fitted models held fixed; nested model variants share the same resampled indices.

%% file: sections/declarations.tex
\section*{Code availability}

The Silscore source code and pretrained model on the main Huesken split are available at 
\url{https://github.com/giusSacco/Silscore}.
The repository includes instructions for using the pretrained model and retraining it on new data,
together with scripts and notebooks to reproduce the analyses reported here, from data processing
and model fitting to benchmarking and figure generation.

\section*{Funding}

G.B. acknowledges support from the Italian Ministry of University and Research (MUR) through the
Fondo Italiano per la Scienza (FIS 3), Grant No. FIS-2024-01745, RNAScale.

\section*{Intellectual property statement}

The Silscore source code and accompanying documentation were deposited with the Benelux Office
for Intellectual Property (BOIP) through its i-DEPOT service on 11 November 2026
(evidence no. 165081; reference SILCOP-001).

\section*{Author contributions}

Giuseppe Sacco, Guido Sanguinetti and Giovanni Bussi conceived and designed the study.
Isabella Baldini, Riccardo Paone and Mattia Capulli contributed to the conception of the study.
Giuseppe Sacco developed the model, implemented the software, curated and processed the
datasets, performed model fitting, statistical analyses, validation and benchmarking, and
prepared the figures.
Guido Sanguinetti and Giovanni Bussi supervised the research.
All authors read and critically revised the manuscript.

%% file: sections/supplementary.tex
\section{Supplementary material}\label{sec:supplementary}

\setcounter{figure}{0}
\renewcommand{\thefigure}{S\arabic{figure}}
\renewcommand{\theHfigure}{S\arabic{figure}}
\setcounter{table}{0}
\renewcommand{\thetable}{S\arabic{table}}
\renewcommand{\theHtable}{S\arabic{table}}

\begin{figure}[!htbp]
\centering
\includegraphics[width=0.95\textwidth]{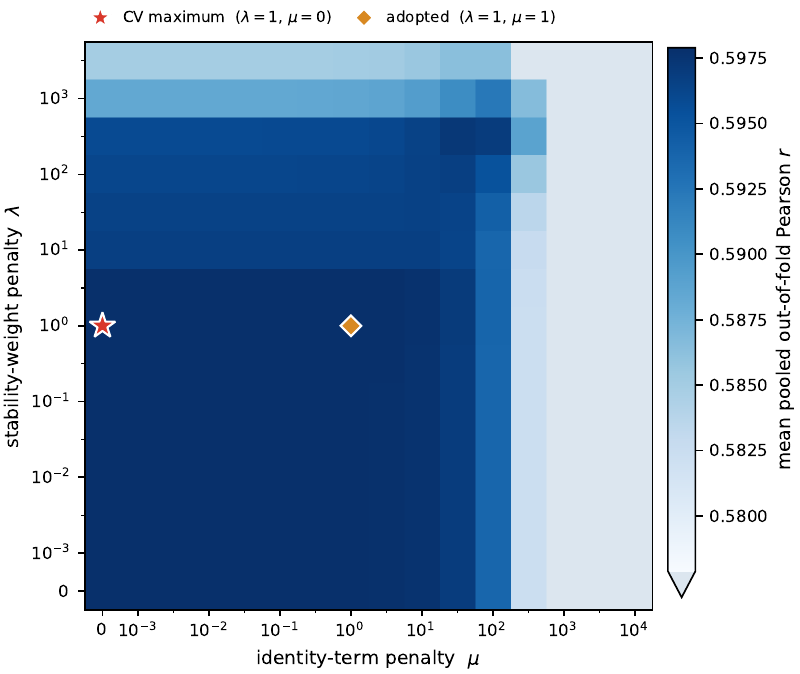}
\caption{Hyperparameter landscape estimated by ten repetitions of five-fold cross-validation
on the Huesken training records. Colours show the mean pooled out-of-fold Pearson correlation.
The star marks the grid maximum and the diamond the adopted pair $(\lambda,\mu)=(1,1)$.
The zero row and column are the unpenalised limits for the corresponding coefficient blocks.
The held-out Huesken records, Mixset and Katoh--Suzuki were not used.}
\label{fig:hparam-landscape}
\end{figure}

\subsection{Sensitivity to identity terms}
\label{sec:global-stability-sensitivity}

The Huesken optimum and penalty strength were stable as identity terms were added. Across
Thermo-only, Thermo+$\mathrm{U1}$ and Silscore, the fitted optimum ranged from $-28.29$ to $-27.21$
kcal/mol and the standardised values of $\kappa$ differed by less than 6\%.

When the linear and quadratic terms in $\dG$ were removed and identity selection repeated, the ablated model
retained 11 identity coefficients rather than six. Five were shared with Silscore; it dropped
$\mathrm{C2}$ and selected six additional terms, giving 29 fitted parameters rather than 26. This is
consistent with identity terms absorbing some of the residual structure otherwise
represented by global stability, although the selection alone does not identify the mechanism.
Despite its three additional parameters, the ablated model gave correlations of $0.640$, $0.561$
and $0.504$ on the held-out Huesken test set, Mixset and Katoh--Suzuki, compared with $0.634$, $0.580$ and
$0.623$ for Silscore. The paired Silscore-minus-ablated differences were $-0.006$
[$-0.045$, $0.036$], $0.019$ [$-0.009$, $0.048$] and $0.120$ [$0.090$, $0.151$], respectively.
Thus the same-dataset held-out evaluation did not expose a substantial loss of robustness under
transfer.

\subsection{Sensitivity of the fitted energy profiles}

\begin{figure}[!htbp]
\centering
\includegraphics[width=0.79\textwidth]{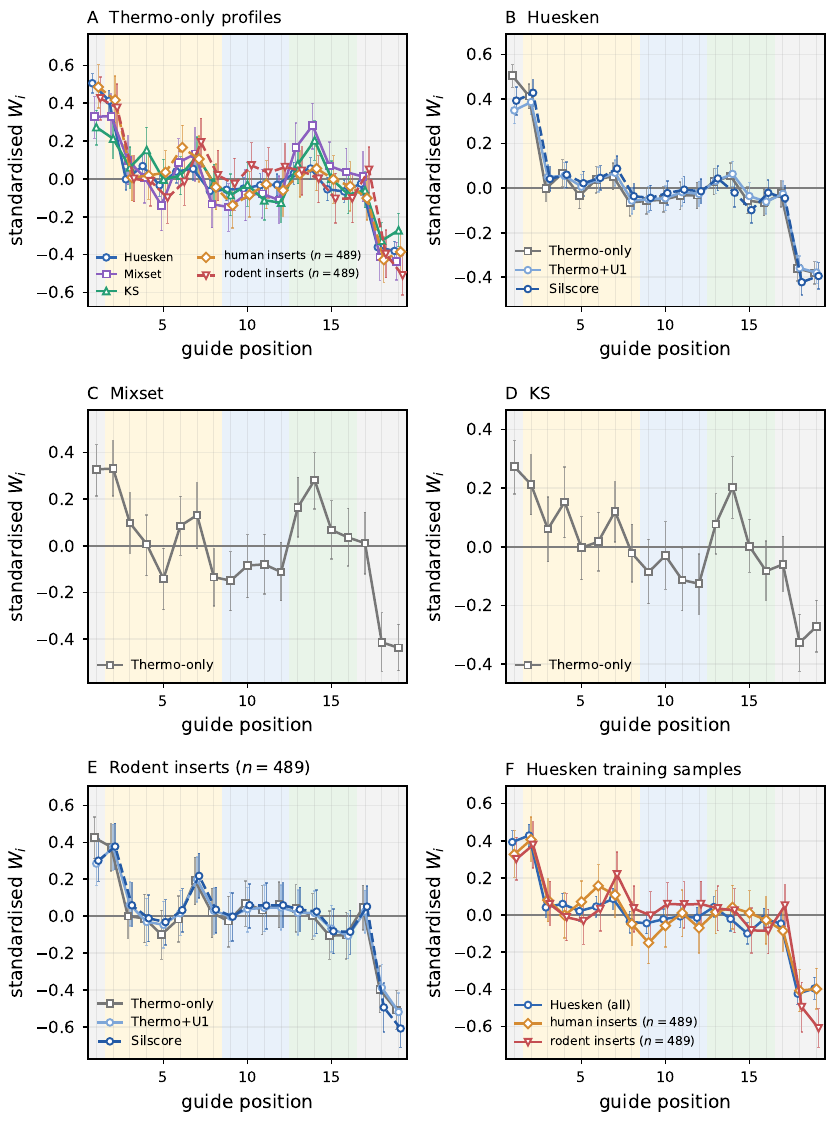}
\caption{Sensitivity of the nucleotide profiles to the fitted model and training sample.
\textbf{A}, Thermo-only profiles fitted independently to all five training samples. \textbf{B--E},
profile sensitivity within Huesken, Mixset, Katoh--Suzuki and the 489-guide rodent-insert training
sample. Thermo-only, Thermo+$\mathrm{U1}$ and Silscore are shown in B and E. In C and D, Silscore selects
no identity term and therefore coincides with Thermo-only; their common profile is shown, while
Thermo+$\mathrm{U1}$ is omitted. \textbf{F}, Silscore profiles for the complete Huesken training set and for the
equally sized human-insert and rodent-insert samples. Values are
standardised by the within-dataset efficacy standard deviation. Error bars are pointwise 95\%
intervals from 4000 bootstrap samples, conditional on the identity terms selected from the
original sample.}
\label{fig:weight-profiles-supp}
\end{figure}

Figure~\ref{fig:weight-profiles-supp} separates the contributions of the physical and identity
terms. The same fitting procedure was used for every profile shown in the figure. In addition to the three datasets in
the main text, we show profiles fitted to two equally sized training samples used in
Section~\ref{sec:targets}: 489 guides against human reporter inserts and 489 against rodent
reporter inserts. The selected identity terms were
$\mathrm{U1}$, $\mathrm{A19}$, $\mathrm{G18}$, $\mathrm{C2}$, $\mathrm{C1}$ and $\mathrm{G14}$ for Huesken, none for Mixset or Katoh--Suzuki, $\mathrm{U1}$ for
the human-insert sample, and $\mathrm{U1}$ plus $\mathrm{G18}$ for the rodent-insert sample.

\begin{figure}[!htbp]
\centering
\includegraphics[width=0.92\textwidth]{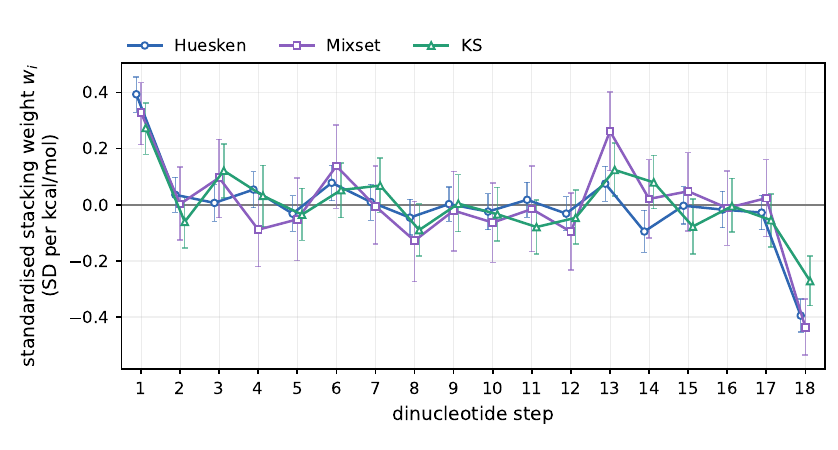}
\caption{Raw Silscore weights $w_i$ of the 18 dinucleotide steps for Huesken, Mixset
and Katoh--Suzuki. These are the fitted coordinates before adjacent steps are summed into the nucleotide
profile $W_i$ shown in Figure~\ref{fig:weight-profiles}. Values are divided by the
within-dataset efficacy standard deviation; error bars are pointwise 95\% bootstrap intervals
from 4000 samples, conditional on the identity terms selected from the original sample.}
\label{fig:weight-stacks-supp}
\end{figure}

\begin{figure}[!htbp]
\centering
\includegraphics[width=0.95\textwidth]{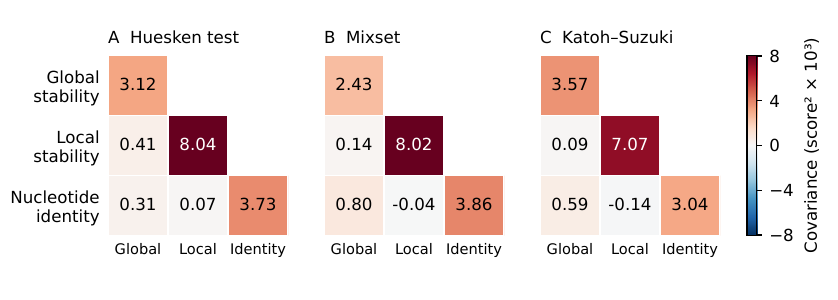}
\caption{Covariance matrices of the three score components on the Huesken test set (249 guides),
Mixset (472) and Katoh--Suzuki (702), using the same model fitted on the 2182 Huesken
training guides without refitting.
Values are in squared score units multiplied by $10^3$;
all three matrices share the same colour scale.}
\label{fig:term-covariance}
\end{figure}
\begin{figure}[!htbp]
\centering
\includegraphics[width=0.86\textwidth]{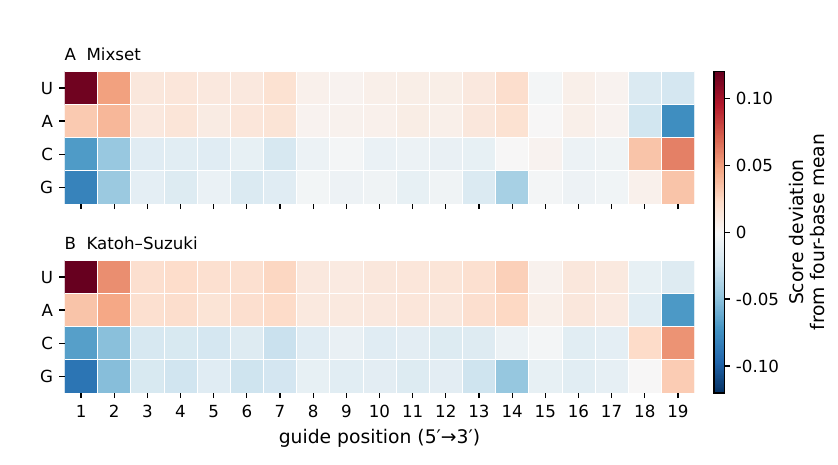}
\caption{As in Figure~\ref{fig:nucleotide-heatmap}, but using the sequences from Mixset
(472 guides) and Katoh--Suzuki (702 guides) with the same Huesken-trained model.
Both maps share the colour scale of the main-text map.}
\label{fig:nucleotide-heatmap-supp}
\end{figure}

\subsection{Comparator implementation}\label{sec:comparators}

We reimplemented DSIR following the published procedure \citep{vertAccurateInterpretableModel2006}.
Because Mixset and Katoh--Suzuki do not report the $3'$ overhang required by the 21-nt representations, sparse-19, spectrum-19 and composite-19/19 were used in every evaluation, whereas composite-21/21 was included only for Huesken.
On the published Huesken split, our implementation reproduced all seven sequence-based test correlations within their published rounding precision.

The re-fitted i-Score comparator uses the architecture of \citet{ichiharaThermodynamicInstabilitySiRNA2007}: 76 position--nucleotide indicators fitted by ridge regression.
The ridge penalty was selected on the 2182 Huesken training records by ten repetitions of five-fold cross-validation over the seven powers of ten from $10^{-3}$ to $10^3$, maximising the Pearson correlation of pooled out-of-fold predictions, averaged across repetitions.
The selected value, $\alpha=10$, was then fixed for every re-fit.
When fitted on the canonical Huesken training set, its 76 coefficients closely reproduced the relative pattern of the published i-Score weights (Pearson $r=0.997$).
It was trained on the same records as Silscore, so this is a comparison with the re-fitted architecture rather than with the published coefficients.
Reynolds and Ui-Tei were evaluated as fixed rule-based scores and were never fitted \citep{reynoldsRationalSiRNADesign2004,ui-teiGuidelinesSelectionHighly2004}.
The sole modification to Reynolds was its hairpin filter: because the original Oligo 6.0 calculation was unavailable, we approximated it using the open seqfold minimum free energy at $37^\circ$C.
Using only the Huesken training sequences, exact agreement with Reynolds scores precomputed in siRNAEfficacyDB \citep{zhangSiRNAEfficacyDBExperimentallySupported2024} showed a plateau around $-0.5$ kcal/mol, so we fixed $\Delta G\geq-0.5$ kcal/mol as the criterion for no stable hairpin across evaluations.
All other criteria followed the published definitions.
Every fitted comparator was re-trained whenever the training set changed.

We used the preprocessing, training and prediction scripts of the latest OligoFormer version
available at the time of writing (commit \texttt{e2f53ad}), without modification.
OligoFormer requires a 57-nt target context: the 19-nt binding site and 19 flanking nucleotides
on each side. Released contexts were used where available; the 70 missing Huesken contexts
were recovered from its original Supplementary Table 2, with boundary padding by \texttt{X}
for nine of these records. Feature extraction followed the released procedure.
Each training pool was divided into 80\% fit and 20\% validation records using five stratified
splits. The evaluation records were excluded from fitting and checkpoint
selection. For each split, five fresh training runs used PyTorch seeds 0--4, set externally before executing
the original entry point; Python and NumPy seeds remained fixed. Model and training settings
were kept at their released defaults, including checkpoint selection and early stopping.
Prediction followed the released procedure with a fixed PyTorch seed.
We report the mean Pearson correlation of the 25 training runs. In each
bootstrap sample, we also independently resample five training--validation splits and five PyTorch seeds
with replacement and average the correlations of their 25 combinations on the sampled guides.